\documentclass[a4paper,11pt]{article}
\usepackage{pos}
\usepackage{soul}
\usepackage{booktabs}
\usepackage{multirow}
\renewcommand{\logo}{\relax}

\title{Dispersive analysis of the $\pi\pi$ and $\pi K$ scattering data}

\author*{Oleksandra Deineka}
\author{Igor Danilkin}
\author{Marc Vanderhaeghen}

\affiliation{Institut f\"ur Kernphysik \& PRISMA$^+$  Cluster of Excellence, Johannes Gutenberg Universit\"at,\\  D-55099 Mainz, Germany}

\emailAdd{deineka@uni-mainz.de}

\abstract{We present a data-driven analysis of the S-wave $\pi\pi \to \pi\pi\,(I=0,2)$ and $\pi K \to \pi K\,(I=1/2, 3/2)$ reactions using the partial-wave dispersion relation. 
The contributions from the left-hand cuts are parametrized using the expansion in a suitably constructed conformal variable, which accounts for its analytical structure. The partial-wave dispersion relation is solved numerically using the $N/D$ method.
The fits to the experimental data supplemented with the constraints from chiral perturbation theory at threshold and Adler zero give the results consistent with Roy-like (Roy-Steiner) analyses. 
For the $\pi\pi$ scattering we present the coupled-channel analysis by including additionally the $K\bar{K}$ channel. 
By the analytic continuation to the complex plane, we found poles associated with the lightest scalar resonances $\sigma/f_0(500)$, $f_0(980)$, and $\kappa/K_0^*(700)$. For all the channels we also performed the fits directly to the Roy-like (Roy-Steiner) solutions in the physical region, in order to minimize the $N/D$ uncertainties in the complex plane and to extract the most constrained Omn\`es functions.}

\FullConference{%
  The 10th International Workshop on Chiral Dynamics - CD2021\\
  15-19 November 2021\\
  Online
  }

\begin{document}
\maketitle

\section{Introduction}

The renewed interest in the hadron spectroscopy has been motivated by recent experimental discoveries of unexpected exotic hadron resonances and the success of lattice QCD, which recently calculated the lowest hadron excitation spectrum with the masses of the light quarks near their physical values. 

To correctly identify resonance parameters, one must search for poles in the complex plane. In order to determine the pole position of the resonance, one has to analytically continue the amplitude to the unphysical Riemann sheets. The proper theoretical framework should satisfy the main principles of the S-matrix theory, namely unitarity, analyticity, and crossing symmetry. These constraints were successfully incorporated in the set of Roy or Roy-Steiner equations. However, the rigorous implementation of these equations requires experimental knowledge of all partial waves with different isospin in the direct and crossing channels (including high energy region). Furthermore, applying Roy-like equations for coupled-channel cases is quite complicated and has not been achieved in the literature so far. Because of the difficulties mentioned above, in the experimental and lattice analyses, it is a common practice to 
rely on simple parameterizations, like superposition of Breit-Wigner resonances or the K-matrix approach. Both methods ignore the existence of the left-hand cut and often lead to spurious poles in the complex plane.

A good alternative to the K-matrix approach and a complementary method to Roy analysis is the so-called $N/D$ technique \cite{Chew:1960iv}, which provides the solution to the dispersion relation for the partial-wave amplitudes. In this method, unitarity and analyticity constraints are implemented exactly. The required input to solve the partial wave dispersion equation is the discontinuity along the left-hand cut, which is typically approximated one way or another using chiral perturbation theory ($\chi$PT). In \cite{Danilkin:2020pak} we extended the ideas of \cite{Gasparyan:2010xz, Danilkin:2010xd, Gasparyan:2011yw, Gasparyan:2012km}, where the left-hand cut contributions were approximated using an expansion in powers of a suitably chosen conformal variable. However, in contrast to \cite{Danilkin:2011fz, Danilkin:2012ap}, we followed a data-driven approach and adjusted the unknown coefficients in the expansion scheme to empirical data (or Roy-like results) directly \cite{Danilkin:2020pak} and used $\chi$PT only for threshold constraints, which can be turned off. In this way, the model dependence is avoided, and the method can also be applied to the reactions which do not include Goldstone bosons (for a recent application to $\gamma\gamma \to D\bar{D}$ scattering, see \cite{Deineka:2021aeu}).

\section{Formalism}

We consider the $2\to 2$ scattering process, which can be described by the $s$-channel partial-wave amplitudes $t^{(J)}_{ab}(s)$, where $ab$ are the coupled-channel indices with $a$ and $b$ standing for the initial and final state, respectively. For the following discussion, we focus only on the S-wave $(J=0)$ and therefore will suppress the label $(J)$. Within the maximal analyticity assumption, the partial-wave amplitudes satisfy the dispersive representation
\begin{equation}\label{DR_0}
t_{ab}(s)=\int_{-\infty}^{s_L}\frac{d s'}{\pi}\frac{\text{Disc } t_{ab}(s')}{s'-s} + \int_{s_{\text{th}}}^{\infty}\frac{d s'}{\pi}\frac{\text{Disc } t_{ab}(s')}{s'-s}\,,
\end{equation}
where $s_{\text{th}}$ being the lowest threshold of the corresponding two-meson system, $s_L$ is the position of the closest left-hand cut singularity and the discontinuity along the right-hand cut is given by the unitarity relation
\begin{align}
    \text{Disc } t_{ab}(s)=\sum_c t_{ac}(s)\,\rho_c(s)\,t_{cd}^{*}(s)\,,
\end{align}
where $\rho_c(s)$ is the phase space factor. The unitarity condition guarantees that the partial-wave amplitudes at infinity approach at most constants. In accordance with that, we can make one subtraction in Eq. (\ref{DR_0}) to suppress the high-energy contribution under the dispersive integrals. Thus we rewrite Eq. (\ref{DR_0}) as
\begin{align}\label{DR_1}
t_{ab}(s)&=t_{ab}(0)+\frac{s}{\pi}\int_{-\infty}^{s_L}\frac{d s'}{s'}\frac{\text{Disc } t_{ab}(s')}{s'-s} + \frac{s}{\pi}\int_{s_{\text{th}}}^{\infty}\frac{d s'}{s'}\frac{\text{Disc } t_{ab}(s')}{s'-s}\nonumber \\
&\equiv U_{ab}(s) + \frac{s}{\pi} \int_{s_{\text{th}}}^{\infty}\frac{d s'}{s'}\frac{\text{Disc } t_{ab}(s')}{s'-s} \,,
\end{align}
where we combined the subtraction constant together with the left-hand cut contributions into the function $U_{ab}(s)$. The solution to (\ref{DR_1}) can be written using the $N/D$ ansatz \cite{Chew:1960iv}
\begin{equation}\label{N/D}
t_{ab}(s)=\sum_c D^{-1}_{ac}(s)\,N_{cb}(s)\,,
\end{equation}
where the contributions of left- and right-hand cuts are separated into $N_{ab}(s)$ and $D_{ab}(s)$ functions, respectively. As a consequence of this ansatz, one needs to solve a system of linear integral equations \cite{Luming:1964zz,Johnson:1979jy}
\begin{align}\label{N-fun}
N_{ab}(s)&=U_{ab}(s)+\frac{s}{\pi} \sum_{c} \int_{s_{\text{th}}}^{\infty}\frac{d s'}{s'}\frac{N_{ac}(s')\,\rho_{c}(s')\,(U_{cb}(s')-U_{cb}(s))}{s'-s}\,\\
\label{D-fun}
D_{ab}(s)&=\delta_{ab}- \frac{s}{\pi} \int_{s_{\text{th}}}^{\infty}\frac{d s'}{s'}\frac{N_{ab}(s')\,\rho_{b}(s')}{s'-s}\,.
\end{align}
with the input of $U_{ab}(s)$ for $s>s_{\text{th}}$. 
Since in a general scattering problem, little is known about the left-hand cuts, except their analytic structure in the complex plane, one can consider an analytic continuation of $U_{ab}(s)$ to the physical region by means of an expansion in a suitably contracted conformal mapping variable $\xi_{ab}(s)$ \cite{Gasparyan:2010xz,Danilkin:2010xd,Gasparyan:2011yw,Gasparyan:2012km}, 
\begin{equation}\label{ConfExpansion}
U_{ab}(s)= \sum_{n=0}^\infty C_{ab,n}\,(\xi_{ab}(s))^n\,,
\end{equation}
which is chosen such that it maps the left-hand cut plane onto the unit circle \cite{Frazer:1961zz}. The form of $\xi_{ab}(s)$ depends on the cut structure of the reaction (i.e. $\{ab\}$) and specified by the position of the closest left-hand cut branching point and an expansion point around which the series is expanded (see Ref.\,\cite{Danilkin:2020pak} for more details). We determine the unknown $C_{ab,n}$ in Eq. (\ref{ConfExpansion}) and the optimal positions of expansion point directly from the fit to the data and use $\chi$PT results only as constraints for the scattering lengths ($a$), slope parameters ($b$), and Adler zero values ($s_A$). In the case of no bound states or CDD poles \cite{Yao:2020bxx,Salas-Bernardez:2020hua}, the Omn\`es function can be easily obtained as
\begin{equation}
    \Omega_{ab}(s)=D_{ab}^{-1}(s)\,.
\end{equation}

\section{Numerical results}
Both $\pi\pi$ and $\pi K$ channels have been measured experimentally (see Refs. \cite{Pelaez:2020gnd, Garcia-Martin:2011iqs} for the overview). However, throughout the whole energy range there are large differences between different data-sets and a careful choice of the data is required to achieve a controllable data-driven description of the phase shifts and inelasticity. In order to be consistent with $\chi$PT in the threshold region, we employ the effective range expansion.
For the $\pi\pi$ and $\pi K$ scattering both $a$ and $b$ have been calculated at NNLO in $\chi$PT \cite{Colangelo:2001df,Bijnens:2004bu}. Since for $\pi K$ scattering the calculation of uncertainties is a bit cumbersome at NNLO and have not been provided in \cite{Bijnens:2004bu}, in our fits we take NNLO $\chi$PT values as central results, but include a conservative error-bar, such that it covers the recent Roy-Steiner results \cite{Pelaez:2020gnd}.
As for the Adler zero, in all numerical fits, we take the NLO result \cite{Gasser:1983yg, Bernard:1990kx, GomezNicola:2001as} as a central value, with the uncertainties from the omitted higher orders as $|\text{NLO}-\text{LO}|$, which should provide a very conservative estimate. The NLO values for the low-energy constants are taken from \cite{Bijnens:2014lea}.  The expansion point, around which the conformal series is expanded, is chosen in the middle between the threshold and the energy of the last data point that is fitted. To assess the systematic uncertainties, we vary this parameter (see Ref. \cite{Danilkin:2020pak} for more details). The uncertainties are propagated using a bootstrap approach.  In several cases, however, we will be fitting Roy (Roy-Steiner) solutions, which are smooth functions and their errors are fully correlated from one point to another. In these cases, $\chi^2/d.o.f$ loses its statistical meaning and can be  $<1$.

All results presented below have been checked to fulfill the partial-wave dispersion relation given in Eq.\,(\ref{DR_1}).

\begin{figure*}[t]
\centering
\includegraphics[width =1\textwidth ]{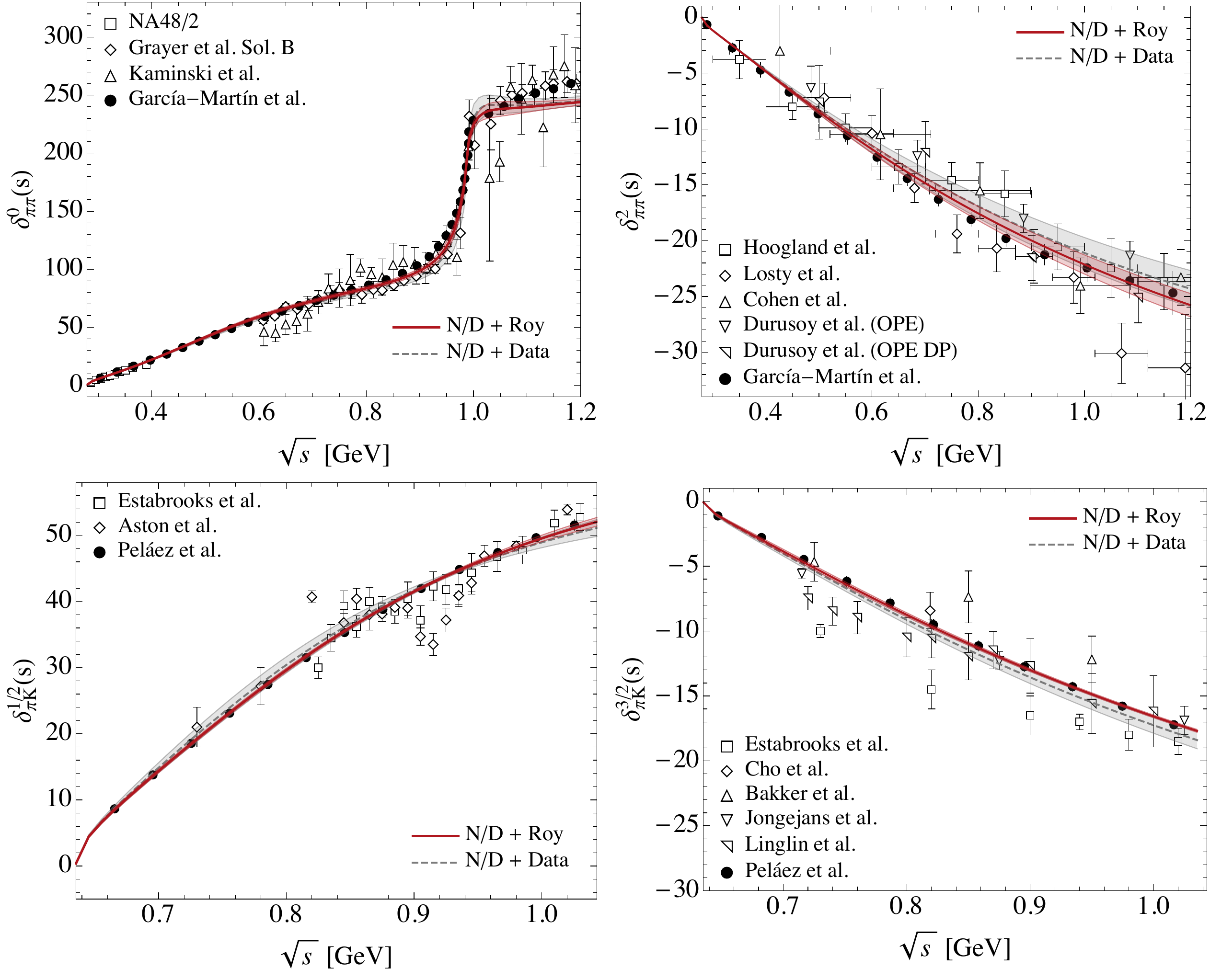}
\caption{Results for the $\pi\pi \to \pi\pi$ and $\pi K \to \pi K$ scattering. Top row corresponds to the $\pi\pi \to\pi\pi$ scattering with $I = 0$ coupled-channel case (left panel) and $I=2$ single-channel case (right panel). Bottom row shows the results for single-channel study of the $\pi K \to\pi K$ scattering with $I = 1/2$ (left panel) and $I=3/2$ (right panel). In the phase shift plots two curves are shown: fit to the experimental data (dashed curve) and fit to the pseudo data from Roy-like analyses \cite{Pelaez:2015qba,Garcia-Martin:2011nna,Garcia-Martin:2011iqs,Pelaez:2019eqa} (thick curve).
}
\label{Fig:pipiSC}
\end{figure*}

\subsection{Analysis of the $\pi\pi$ data for $I=0,2$}

\begin{figure*}[t]
\centering
\includegraphics[width =1\textwidth ]{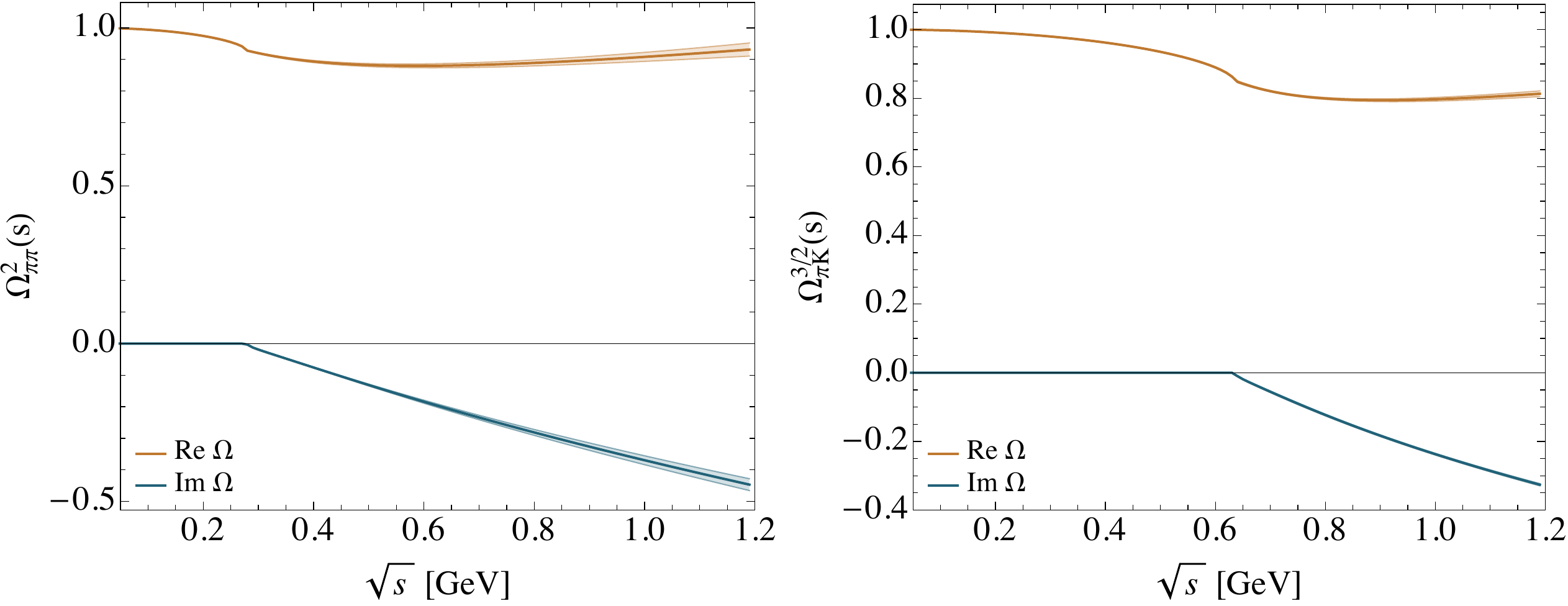}
\caption{Omn\'es function for $\pi\pi \to \pi\pi$ ($I=2$) (left panel) and $\pi K \to \pi K$ ($I=3/2$) (right panel) processes.
}
\label{Fig:Omnes}
\end{figure*}

For the isoscalar $\pi\pi$ scattering, already the single channel analysis provides a realistic estimate of the resonance position of $\sigma/f_0(500)$, which is known to be connected almost exclusively to the pion sector. 
However, a comprehensive study of the region up to $\sqrt{s}=1.2$ GeV should account for the interplay between $\pi\pi$ and $K\bar{K}$ channels. 
In the physical region the two-channel $t$-matrix is fully described by experimental information on the $\pi\pi$ phase shift $\delta_{\pi\pi}(s)$, the inelasticity $\eta(s)$ (or $|t_{\pi\pi,K\bar{K}}(s)|$ for $s>4m_K^2$) and the $\pi\pi \to K\bar{K}$ phase $\delta_{\pi\pi,K\bar{K}}(s)$. We first fit the available experimental data supplemented with constraints for scattering length, slope parameter and Adler zero from $\chi$PT in the $\pi\pi \to \pi\pi$ channel. As for the $\pi\pi\to K \bar{K}$ channel, the  complication stems from two facts. Firstly, the experimental data exist only in the physical region above $K\bar{K}$ threshold. Therefore, in order to stabilize the fits, we make sure that the obtained $|t_{\pi\pi,K\bar{K}}(s)|$ stays small around $s=0$ as a manifestation of $\chi$PT. Secondly, the existing experimental data for both $|t_{\pi\pi,K\bar{K}}(s)|$ and $\delta_{\pi\pi,K\bar{K}}(s)$ contains incompatible data sets and require to make some choices \cite{Danilkin:2020pak}. The best fit leads to $\sqrt{s_\sigma}= 454(12)^{+6}_{-7} - 262(12)^{+8}_{-12}\,i$ MeV and $\sqrt{s_{f_0}}= 990(7)^{+2}_{-4} - 17(7)^{+4}_{-1}\, i$ MeV. On the other side, we have at our disposal very precise $\pi\pi \to \pi\pi$ Roy-like analyses from \cite{Garcia-Martin:2011nna, Garcia-Martin:2011iqs, Pelaez:2019eqa} and $\pi\pi \to K\bar{K}$ Roy-Steiner analyses from  \cite{Buettiker:2003pp, Descotes-Genon:2006sdr,Pelaez:2020uiw,Pelaez:2020gnd, Pelaez:2018qny}. Unfortunately, they do not come from the coupled-channel Roy-Steiner analyses and may display some inconsistencies between each other in the two channel approximation. In order to avoid possible conflict, we impose the $\pi\pi \to K\bar{K}$ Roy-Steiner solution only as constraint on $|t_{\pi\pi,K\bar{K}}(s)|$ in the unphysical region $4m_\pi^2<s<4m_K^2$. As for the $\delta_{\pi\pi,K\bar{K}}(s)$, we take advantage of experimental data of Cohen et al. \cite{Cohen:1980cq} in the fit, which are quite precise. The good description of the data is shown in Fig. \ref{Fig:pipiSC} (top left panel). The resulting pole positions come relatively close to the current Roy-like analyses average results:
\begin{align}
    \sqrt{s^{N/D}_\sigma}&= 458(10)^{+7}_{-15}- i\,256(9)^{+5}_{-8} \text{ MeV \cite{Danilkin:2020pak}}\,, &\sqrt{s^{\text{Roy}}_\sigma}&=449^{+22}_{-16}-i\,275(15) \text{ MeV \cite{Pelaez:2015qba, Pelaez:2021dak}}\, ,\nonumber\\ 
   \sqrt{s^{N/D}_{f_0}}&=993(2)^{+2}_{-1} - i\,21(3)^{+2}_{-4} \text{ MeV \cite{Danilkin:2020pak}}\,, &\sqrt{s^{\text{Roy}}_{f_0}}&=996^{+7}_{-14}-i\,25^{+11}_{-6} \text{ MeV \cite{Pelaez:2019eqa,Garcia-Martin:2011nna,Garcia-Martin:2011iqs,Moussallam:2011zg}}\,.\nonumber
\end{align}
We now turn to the partial wave dispersion relation analysis of the non-resonant $\pi\pi\,(I=2)$ scattering, which was not given in \cite{Danilkin:2020pak}. While the overall strategy remains the same as for the $I=0$ case, we notice that by fitting either experimental data or the Roy analysis results, supplemented with $\chi$PT constraints, we obtain an unphysical zero of $D(s)$ far away from the threshold on the first Riemann sheet. To avoid this artificial bound state, we impose in the fit the exact fulfilment of p.w. dispersion relation given by Eq.\,(\ref{DR_1}). This leads to the four-parameter fit with a good description of both the experimental data and the Roy-like analysis results. The phase shift obtained from this fit is shown in Fig. \ref{Fig:pipiSC} (top right panel). The values of the fitted parameters, threshold parameters and Adler zeros are collected in Tables \ref{tab:FitResults} and \ref{tab:ThresholdParameters}. The Omn\`es function is shown in Fig. \ref{Fig:Omnes}.

\subsection{Analysis of the $\pi K$ data for $I=1/2,3/2$}

The available experimental data for these processes is scarce in the region close to the $\pi K$ threshold, and contain inconsistencies even within one dataset \cite{Estabrooks:1977xe, Bakker:1970wg}.
 Therefore the fits are strongly affected by the $\chi$PT low energy constraints. The most precise calculation of the scattering length and slope parameter in $\chi$PT has been performed at NNLO in \cite{Bijnens:2004bu}. While the result for the scattering length is consistent with the recent Roy-Steiner predictions, it seems that there is a small tension in the slope parameter value. Therefore in our fits to the experimental data (or pseudo-data from Roy-Steiner analyses) we take NNLO $\chi$PT values as central results, but include the conservative error-bar, such that it covers the recent Roy-Steiner results \cite{Pelaez:2020uiw, Pelaez:2020gnd}. As for the Adler zeros, we take the NLO values. Since we consider only the single-channel approximation for $I=1/2$ (and $I=3/2$), we perform the fit till $\eta K$ threshold of the experimental data. 
 In this way we also exclude the influence of the $K_0^*(1430)$ resonance.  

For $I=1/2$ we observed in \cite{Danilkin:2020pak}
that fitting the experimental data 
or Roy-Steiner analysis of \cite{Pelaez:2020uiw, Pelaez:2020gnd} provided equivalent four parameter fits with close $\kappa/K_0^*(700)$ pole positions. 
In general, these results compare well with the Roy-Steiner pole position
\begin{align}
    \sqrt{s^{N/D}_\kappa}&= 702(12)^{+4}_{-5}-i\,285(16)^{+8}_{-13} \text{ MeV \cite{Danilkin:2020pak}}\,, &\sqrt{s^{\text{Roy-Steiner}}_\kappa}&=653^{+18}_{-12}-i\,280(16) \text{ MeV } \,.\nonumber
\end{align}
The latter we took as a conservative average between \cite{Buettiker:2003pp, Descotes-Genon:2006sdr} and \cite{Pelaez:2020uiw,Pelaez:2020gnd}.
The one-sigma difference in the resonance mass can be attributed to the fact, that we are fitting Roy-Steiner solution only in the elastic region. The resulting phase shifts are shown in Fig \ref{Fig:pipiSC}.

The situation for the non-resonant $I=3/2$, $\pi K$ scattering resembles the $I=2$, $\pi\pi$ scattering. Again, by fitting either experimental or Roy-Steiner pseudo-data we obtain an unphysical bound state far away from the threshold. Hence, we impose in the fit the exact fulfilment of p.w. dispersion relation given by Eq.\,(\ref{DR_1}) as an additional constraint and obtain the four-parameter fits. 
The fit to experimental data give us the result consistent with Roy-Steiner analysis  \cite{Pelaez:2020uiw,Pelaez:2020gnd} including the slope parameter, which tends towards the value $ m_\pi^3\,b^{\text{Roy-Steiner}}=
-0.0471(49)$ (see Table\,\ref{tab:ThresholdParameters}). To minimize the uncertainties we also fitted directly the pseudo-data from Roy-Steiner analysis.
The phase shifts obtained from these fits are shown in Fig.\,\ref{Fig:pipiSC}. The values of the fitted parameters, threshold parameters and Adler zeros are collected in Tables \ref{tab:FitResults} and \ref{tab:ThresholdParameters}. The Omn\`es function is shown in Fig.\,\ref{Fig:Omnes}.

\begin{table}[]
\centering
\begin{tabular}{@{}lcccccc@{}}
\toprule\toprule
\multicolumn{1}{c}{} & $\sqrt{s_E}$, MeV    & $C_0$      & $C_1$      & $C_2$     & $C_3$      & $\chi^2/\text{d.o.f}$ \\ \midrule
\multicolumn{7}{l}{$\pi\pi\to\pi\pi, I = 2$}                                                                           \\ \midrule
Fit to Exp.                 & \multirow{2}{*}{740} & $-16.9(1.1)$ & $-27.1(2.7)$ & $30.2(9.3)$ & $48.3(13.5)$ & 0.96                  \\
Fit to Roy                  &                      & $-18.0(7)$   & $-30.1(1.9)$ & $27.4(5.6)$ & $49.3(8.1)$  & 0.38                  \\ \midrule
\multicolumn{7}{l}{$\pi K\to\pi K, I = 3/2$}                                                                           \\ \midrule
Fit to Exp.                  & \multirow{2}{*}{818} &  $-20.7(8)$  &  $34.8(3.0)$ & $25.7(8.0)$ & $-52.0(11.3)$ & 5.64                    \\
Fit to Roy                  &                      &  $-19.8(2)$  &  $34.0(1.6)$ & $21.1(1.9)$ & $-46.4(4.5)$  & 0.21                      \\ \bottomrule\bottomrule
\end{tabular}
\caption{Fit parameters entering Eq. (\ref{ConfExpansion}) which were adjusted to reproduce the experimental data (denoted Exp.) and the most recent Roy-like results (Roy). The parameters for the other channels can be found in \cite{Danilkin:2020pak}.}
\label{tab:FitResults}
\end{table}

\begin{table}[]
\centering
\begin{tabular}{@{}lcccccc@{}}
\toprule\toprule
\multicolumn{1}{c}{} & $\sqrt{s_A}$, MeV & $m_\pi a$ & $m_\pi^3 b$                    & $\sqrt{s^{\text{NLO}}_A}$, MeV        & $m_\pi a^{\text{NNLO}}$                  & $m_\pi^3 b^{\text{NNLO}}$                \\ \midrule
\multicolumn{7}{l}{$\pi\pi\to\pi\pi, I = 2$}                                                                                                                               \\ \midrule
Fit to Exp.                 & $183(13)$           & $-0.044(1)$ & \multicolumn{1}{c|}{$-0.080(1)$} & \multirow{2}{*}{182(15)} & \multirow{2}{*}{$-0.044(1)$} & \multirow{2}{*}{$-0.080(1)$} \\
Fit to Roy                  & $180(12)$           & $-0.044(1)$ & \multicolumn{1}{c|}{$-0.081(1)$} &                          &                            &                            \\ \midrule
\multicolumn{7}{l}{$\pi K\to\pi K, I = 3/2$}                                                                                                                               \\ \midrule
Fit to Exp.                  & $522(11)$      &   $-0.048(5)$  & \multicolumn{1}{c|}{$-0.056(4)$}          & \multirow{2}{*}{526(11)} & \multirow{2}{*}{$-0.047$}    & \multirow{2}{*}{$-0.027$}    \\
Fit to Roy                  &  $524(9)$     &  $-0.047(4)$     & \multicolumn{1}{c|}{$-0.053(1)$}          &                          &                            &                            \\ \bottomrule\bottomrule
\end{tabular}
\caption{Fit results for the threshold parameters $a$ and $b$ and the Adler zeros $s_A$ (left columns) compared to $\chi$PT values (right columns). The uncertainties on NLO Adler zero positions we estimated as $|\text{NLO}-\text{LO}|$, as explained in the text. The NNLO $\chi$PT values for the threshold parameters are taken from \cite{Bijnens:2004bu,Colangelo:2001df}. The parameters for the other channels can be found in \cite{Danilkin:2020pak}.}
\label{tab:ThresholdParameters}
\end{table}

\section{Conclusion and outlook}

We presented a data-driven analysis of the S-wave $\pi\pi \to \pi\pi$ $(I=0,2)$ and $\pi K \to \pi K$ $(I=1/2,3/2)$ reactions using the p.w. dispersion relation. In this approach unitarity and analyticity constraints are implemented exactly. We accounted for the contributions from the left-hand cuts using the expansion in a conformal variable, which maps the left-hand cut plane onto the unit circle. Then, the once subtracted p.w. dispersion relation was solved numerically by means of the $N/D$ method.

Using existing experimental information and constraints from $\chi$PT at very low energies we applied the p.w. dispersion relation for $I=0,1/2$, where the positions of $\sigma/f_0(500)$, $f_0(980)$ and $\kappa/K_0^*(700)$ have already been obtained from the sophisticated Roy and Roy-Steiner analyses. We demonstrated that the results for the pole parameters are stable and almost do not change if we replace the existing experimental data with the very precise pseudo-data generated by Roy and Roy-Steiner solutions in the physical region. For the non-resonance channels with $I=2$ and $I=3/2$, the fits to experimental data (supplemented with $\chi$PT constraints at threshold and Adler zero) give the results very close to Roy-like solutions. In particular, it is interesting for the $I=3/2$ channel, where there is a tension between the slope parameter calculated in $\chi$PT at NNLO and recent Roy-Steiner extraction.

The proposed data-driven method is not only limited to the $\pi\pi$ and $\pi K$ scattering \cite{Danilkin:2020pak}. Recently, we applied it to the $\gamma\gamma \to D\bar{D}$ reaction, where based on the experimental information we found a bound state just below the $D\bar{D}$ threshold \cite{Deineka:2021aeu}.

\bibliographystyle{JHEP}
\bibliography{bibliography}

\end{document}